\documentclass[fleqn,usenatbib]{mnras}

\usepackage{newtxtext,newtxmath}

\usepackage[T1]{fontenc}
\usepackage{ae,aecompl}

\usepackage{graphicx}	
\usepackage{amsmath}	
\usepackage{amssymb}	
\usepackage{bm}

\title[Tiny Star Clusters]{Nine tiny star clusters in
  Gaia DR1, PS1 and DES}

\author[G.~Torrealba et al.]{
G.~Torrealba,$^{1}$\thanks{E-mail: gtorrealba@asiaa.sinica.edu.tw}
V.~Belokurov,$^{2,3}$
S.~E.~Koposov$^{4,2}$
\\
$^{1}$ Institute of Astronomy and Astrophysics, Academia Sinica, P.O. Box 23-141, Taipei 10617,Taiwan\\
$^{2}$ Institute of Astronomy, University of Cambridge, Madingley Road, Cambridge CB3 0HA, UK\\
$^{3}$ Center for Computational Astrophysics, Flatiron Institute, 162 5th Avenue, New York, NY 10010, USA\\
$^{4}$ McWilliams Center for Cosmology, Department of Physics, Carnegie Mellon University, 5000 Forbes Avenue, Pittsburgh, PA 15213, USA
}

\date{Accepted XXX. Received YYY; in original form ZZZ}

\pubyear{2018}

\begin{document}
\label{firstpage}
\pagerange{\pageref{firstpage}--\pageref{lastpage}}
\maketitle

\begin{abstract}

We present the results of a systematic Milky Way satellite search
performed across an array of publicly available wide-area photometric
surveys. Our aim is to complement previous searches by widening the
parameter space covered. Specifically, we focus on objects smaller
than $1\arcmin$ and include old, young, metal poor and metal rich
stellar population masks. As a result we find 9 new likely genuine
stellar systems in data from GAIA, DES, and Pan-STARRS, which were
picked from the candidate list because of conspicuous counterparts in
the cut-out images. The presented systems are all very compact
($r_h<1\arcmin$) and faint ($M_V\gtrsim-3$), and are associated either
with the Galactic disk, or the Magellanic Clouds.  While most of the
stellar systems look like Open Clusters, their exact classification
is, as of today, unclear. With these discoveries, we extend the
parameter space occupied by star clusters to sizes and luminosities
previously unexplored and demonstrate that rather than two distinct
classes of Globular and Open clusters, there appears to be a continuity
of objects, unmarked by a clear decision boundary.
\end{abstract}

\begin{keywords}
Galaxy: halo, galaxies: dwarf, Magellanic Clouds
\end{keywords}



\section{Introduction}

Is there a sea of undiscovered faint satellites out there in the halo
of the Galaxy? Chances are, there is, given the avalanche of recent
discoveries \citep[see e.g.][]{Koposov2015, 2015ApJ...807...50B} and
the theoretical expectations \citep[][]{Koposov2008, Tollerud2008,
  Koposov2009, Jethwa2018}. With the data from ambitious all-sky
surveys and smaller targeted campaigns, dwarf galaxies and star
clusters are now identified routinely at levels of surface brightness
previously unimaginable \cite[for reviews, see][]{Willman2010,
  Belokurov2013}. The extension of the spectrum of Galactic fragments
into such low luminosity regime is not only a compelling evidence for
the hierarchical structure assembly, but also a convenient benchmark
of the (currently poorly constrained) star-formation theory \citep[see
  e.g.][]{Koposov2009, Penarrubia2012, Brooks2014}.

Star clusters (like those presented in this work) are the faintest
(and the smallest) of the newly identified satellites
\citep[e.g.][]{Belokurov2014, Laevens2014,
  Kim2015a,Kim2015b,2017MNRAS.470.2702K,2017MNRAS.468...97L,2018MNRAS.478.2006L}. Not
only are these hard to find, they are also notoriously problematic to
characterize, as their inner regions suffer from heavy blending if
observed from the ground. While, in principle, there should be a
continuous distribution of cluster properties, it is customary to
divide them into two groups: Globular and Open (GCs and OCs). GCs are
denser, older and more luminous, OCs are more sparse, contain fewer
stars and are substantially younger.

\begin{figure*}
    \includegraphics[width=\textwidth]{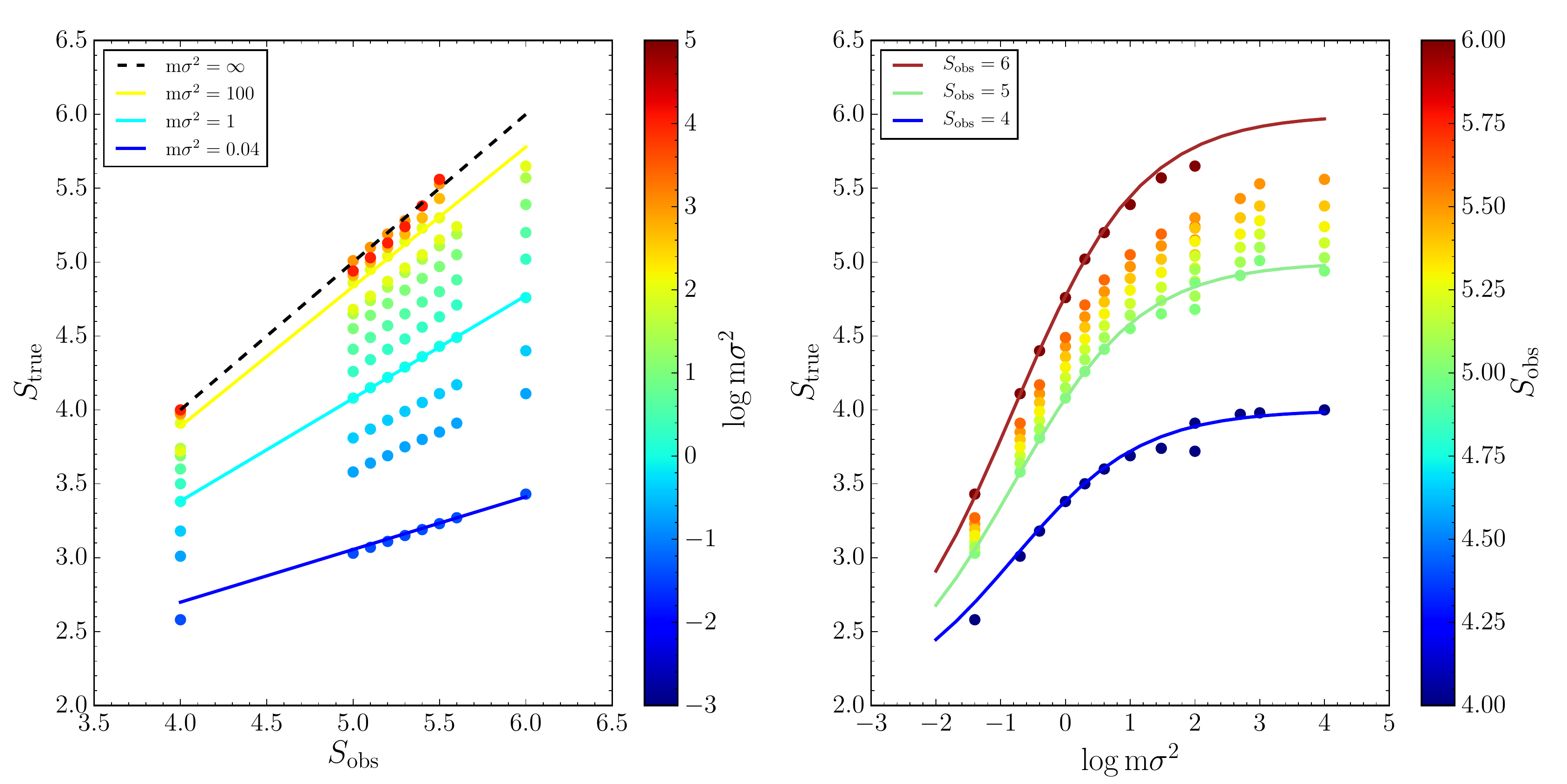}
    \caption{ Comparison between the real significance and the measured
    significance is shown in the left panel for the Poisson noise simulations. The colored lines shows the empirical fit given by equation \ref{eq:Sig2} with the 
    logarithm of the stars per kernel shown in color. The black dashed line is the 1 to 1
    line. Note the evident difference between the real and measured
    significances for low number of stars and how the correction is able to
    recover the real significance from the measured significance. In the right
    panel, we show the real significance as a function of stars per kernel
    for different measured significances. The lines shows the real significance
    calculated from the measured significance.}
    \label{fig:Significance2}
\end{figure*}

There also appears to be a dichotomy as to the habitat of each species. In general, OCs are located around the Galactic plane while GCs wander in the halo. Such difference in spatial distribution betrays the difference in the
birth place. OCs have presumably been born in the Milky Way's disk while the
majority of GCs must have been accreted and their former hosts destroyed
\citep[see e.g.][]{Zinn1993,MackeyGilmore2004,Mackey2010}. In particular, a
sub-population of the halo GCs with intermediate ages (also known as young
halo clusters) appear to have properties indistinguishable from those clusters
found in external galaxies \citep[][]{MackeyGilmore2004}. Even if a direct
comparison is not possible, properties of the stellar populations give away
the GCs' origin, as it is the case of a group of stray halo clusters which follow
the same age-metallicity relationship as that of the stars in the Sgr dwarf
galaxy \citep[see e.g.][]{Marin-Franch2009}.

Importantly, the halo GCs possess the key information as to the
accretion history of the Galaxy: their ages can be used to date the
past merger events. This sort of evidence is vital for dynamical
modelling of the halo assembly, but not directly available otherwise
\citep[see, however,][]{deBoer2015}. Intriguingly, detailed studies of
many of the recently discovered faint halo star clusters confirm that
these too are significantly younger than the quintessential old halo
GCs. For example, Whiting 1 is estimated to be 5 Gyr old
\citep[][]{Carraro2005}, Gaia 1 and 2, 6 and 8 Gyr correspondingly
\citep[][]{2017MNRAS.470.2702K}, even Segue 3, originally deemed to be
truly old \citep[][]{Fadely2011}, is claimed to have an age of $\sim$3
Gyr \citep[][]{Ortolani2013}. With extremely low stellar densities and
(allegedly) young ages, these newly discovered clusters appear to look
more like OCs than GCs.

A deluge of star clusters is now entering the Milky Way, dragged in with their
current host, the Large Magellanic Cloud (LMC). The LMC boasts an impressive
array of star clusters: the early catalogue of \citet{Kontizas1990} lists 1762
candidates, while \citet{Bica2008} announces a record number of 3740 clusters,
but even today new clusters are being added
\citep[see][]{2017AcA....67..363S}, which suggests that the star cluster
census is still incomplete. Furthermore, only a fraction (687) have well-
measured structural parameters \citep[][]{Werchan2011}. The LMC's rich
collection of star clusters has long been used to good effect to forward our
understanding of emergence and development of structure on small scales. This
is illustrated by studies of star cluster dynamical evolution in general
\citep[see e.g.][]{Elson1987}, and under tidal stresses in particular
\citep[][]{Bica2015}. Detailed analysis of the LMC's star-formation history
was possible thanks to the availability of a large number of star clusters
with ages and metaillicities \citep[see e.g.][]{Girardi1995, Harris2004,
Glatt2010,   Palma2015}.  Additionaly, star clusters are perfect tracer
particles to probe the structural properties of the Clouds themselves
\citep[see e.g.][for the latest effort]{Bica2008}.

As samples of new faint objects continue to grow rapidly, selection
biases become more apparent. Almost all satellites discovered in this
century have been found as stellar over-densities in large photometric
datasets. These searches typically start by creating density map of
stars, and convolving them with different kernels to estimate the
local density of stars and the local background density. Then, by
comparing the two densities, one can compute a statistical
significance, which is then used to create a list of candidates
\citep[see][]{Irwin1994}. Evidently, at low levels of statistical
significance, these algorithms begin to incur a sharply increased rate
of false positives. To keep the satellite selection contaminant-free
and to avoid as much of visual inspection of candidates as possible,
it is customary to accept objects at the significance levels higher
than dictated by the probability of a chance fluctuation. More
worryingly, the estimate of the significance itself can be biased, for
example in the case of a small number of stars per kernel or in the
presence of galaxy contamination.

\begin{figure}
    \includegraphics[width=0.98\columnwidth]{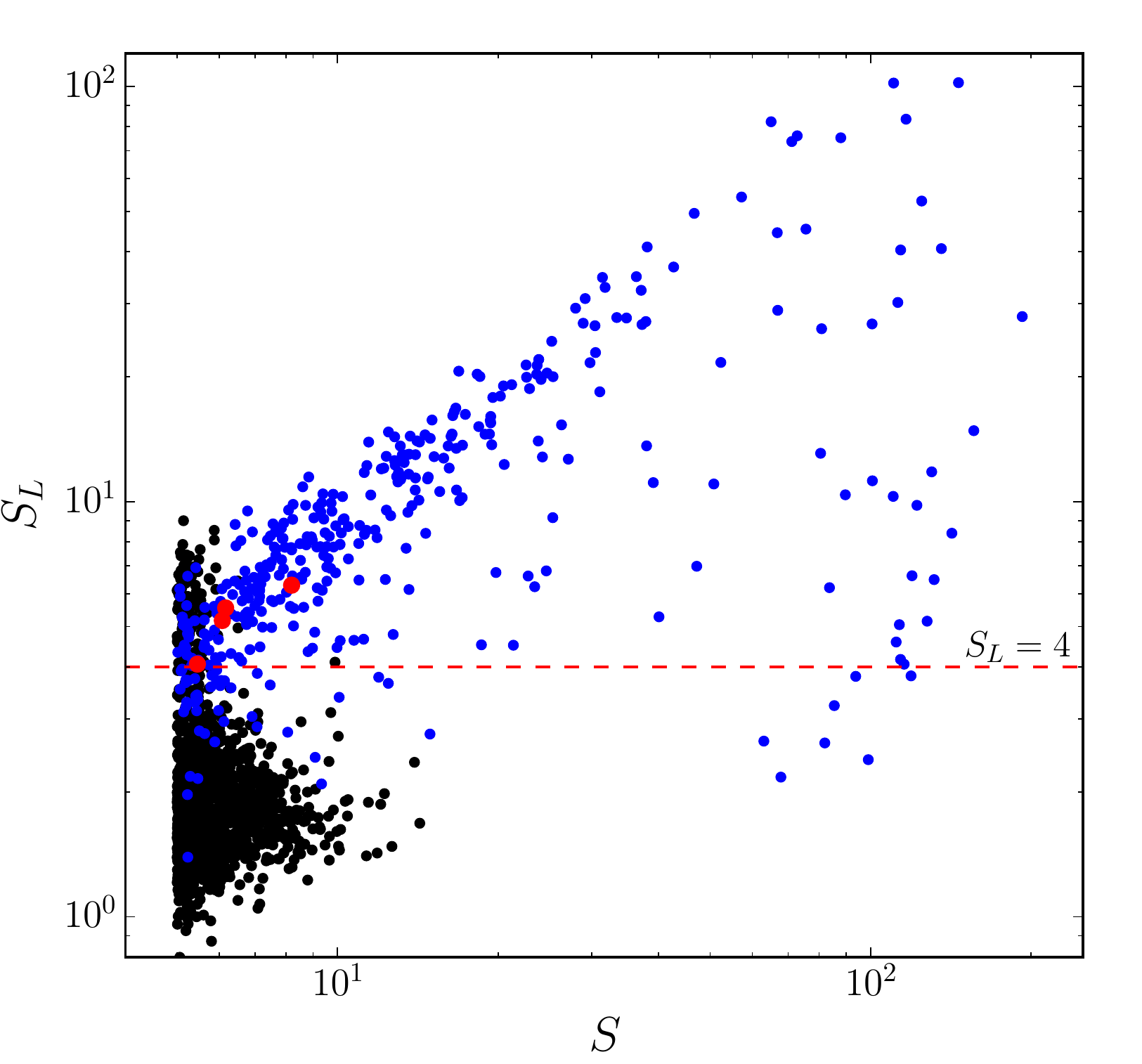}
    \caption{Significance verus local significance for detections in
      Gaia with $1\arcmin$ kernel. Candidates are shown in black,
      previously known MW stellar systems are shown in blue, and new
      discoveries are shown in red. The objects below the $S_L=4$ line
      are excluded to produce a cleaner candidate list. This cut is
      largely supported from the locus that known objects occupy in
      the plot, which is shared by the new discoveries.}
    \label{fig:sslchap2}
\end{figure}

Within a wide range of distances, sizes and luminosities, both globular
clusters and dwarf galaxies can reveal themselves as a compact excess of stars
above the smoothly-varying Galactic density background. Indeed, if a satellite
is resolved into stars, our satellite searching method can be employed to pinpoint
its location in the sky. It would be preferable if such a catalogue-based
search depended as little as possible on the satellite's structural parameters
and/or its stellar populations properties. Therefore, in our approach, we
first cast the net as wide as possible to pick out as many statistically
significant stellar over-densities as possible. The result of the first step
of our algorithm is a list of over-density candidates ranked by their
significance. As a second step, we construct a set of test statistics to
further assess the reality of the pre-selected candidates. The most relevant
of these involves the modeling of each individual candidate to determine
whether the distribution of their stars in the sky and magnitude space,
resembles that of a genuine satellite (i.e. an agglomeration of stars evolving
together at the same Galacto-centric distance).

In this work we choose to conduct a satellite search using a version
of the well-established satellite search algorithm
\citep[see][]{Torrealba2016,2018MNRAS.475.5085T} to look for
small/compact objects ($r<1\arcmin$) in the publicly available data
from different surveys. Several groups (including ourselves) have
already combed these data for the most obvious new objects. However,
the question of robust detection of systems at the low significance
end has not been addressed in detail. Additionally, no extensive
search for small size satellites has been carried out. As a result of
the search, we found 9 new stellar systems across the sky.

\section{All surveys search}	

\begin{figure*}
    \includegraphics[width=\textwidth]{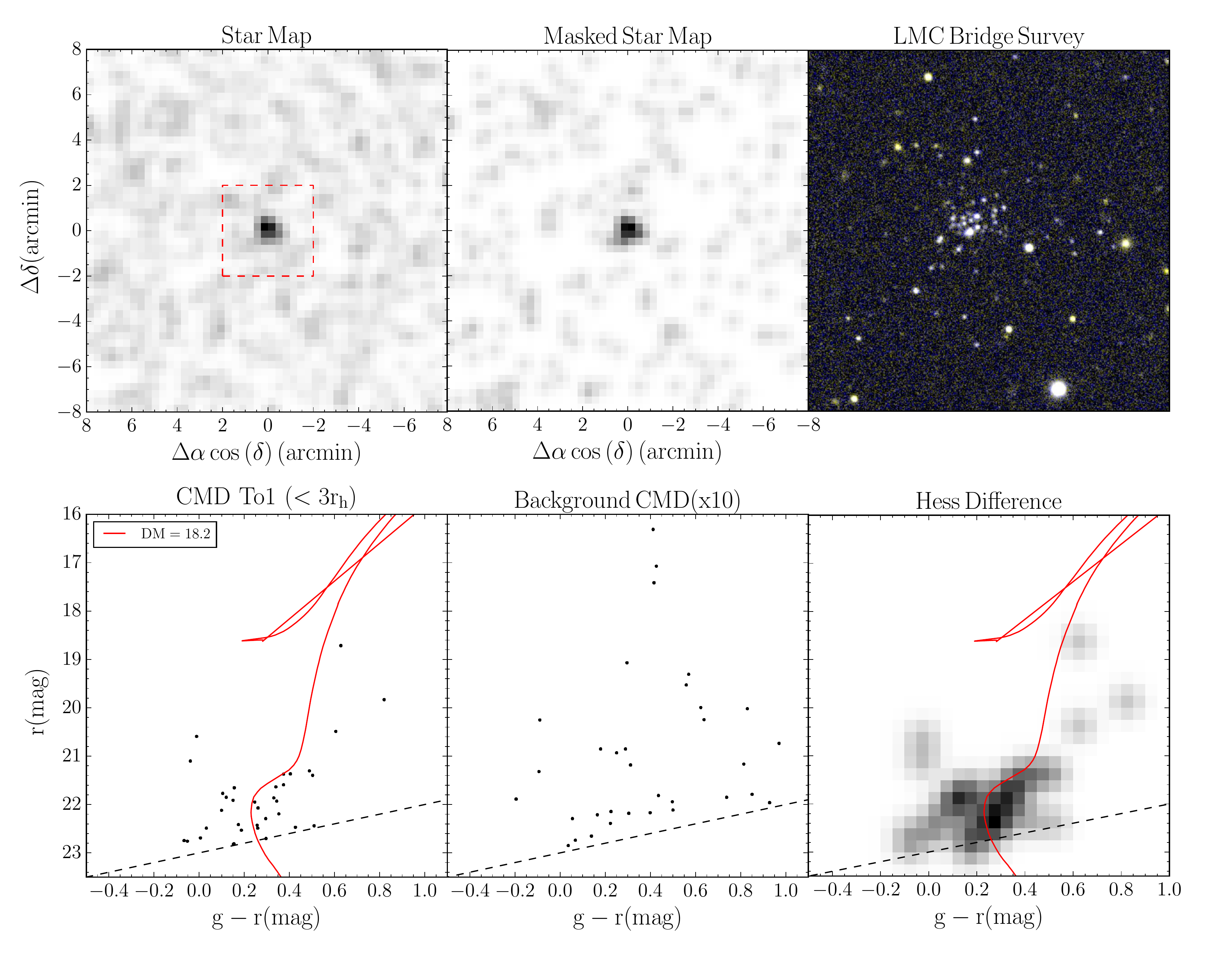}
    \caption{Discovery plot of To~1 in MC bridge survey data. In the top row, left panel shows a density map of the stars in the region, middle panel shows a density map of stars within the isochrone mask of the best fit isochrone, and the rightmost panel shows a false color image of the region marked with the red dashed square in right panel. In the bottom row, left panel shows the CMD of the stars within 3 half light radius of the best fit model, and the best fit isochrone in red. The dashed black line shows the approximate magnitude limit of the survey in this region. The middle lower panel shows the CMD of the background/foreground stellar population for ten times the area used in the lower left panel, the right lower panel shows the Hess difference diagram. The CMD shows a very clear turn off and a sparsely populated red giant branch, in a region where the expected background is scarce. Indeed, it is so sparse that in order to see it, we used an area 10 times bigger than the area used for the satellite. The compactness and luminosity of To~1 are consistent with a faint globular cluster at $\sim43.6$ kpc. Located about $\sim12$ kpc from the LMC, it is highly likely part of its globular cluster system.}
    \label{fig:To1}
\end{figure*}

Most wide-area optical surveys have already been mined for satellites,
but typically aiming at old metal poor populations
\citep[e.g.][]{Koposov2015}. Moreover, searches for objects with sizes
smaller than $1\arcmin$ are often avoided as they are computationally
more expensive and/or might be severely biased if small number
statistics are not taken into account.  In this work, and as a way to
complement previous searches, we have run a search for small clusters
in a set of publicly available surveys. The method used closely
follows the one presented in \citet{2018MNRAS.475.5085T}, in which we
first filter the desired stellar population, create a density map of
stars and convolve it with two kernels: an inner kernel to estimate
the local density, and an outer kernel to estimate the background. We
then estimate two significances, the ``true'' significance $S$, in
which we compare the local density with the expected variance, and a
local significance $S_L$, which helps to identify problematic areas in
which the variance is underestimated. $S_L$ is defined based on the
properties of $S$ around an overdensity:

\begin{equation}
  S_L=\frac{S(0)-\left<S_{d<\sigma_o}\right>}{\sqrt{\rm{Var}\left(S_{d<\sigma_o}\right)}},
\end{equation}

\noindent where $S(0)$ is the significance at the center of an overdensity,
$\sigma_o$ is the width of the outer kernel, and $S_{d<\sigma_o}$ are the
significances for all pixels within $\sigma_o$ from the center. In areas where
the variance is underestimated, $S$ is overestimated and then
$\left<S_{d<\sigma_o}\right>\gg 0$, which means $S_L\ll S$. This allows us to
cull false positives by simply selecting overdensities that have both large
$S$ and large $S_L$.

Specifically, we do a search with two very small inner kernels
($0.3\arcmin$ and $1\arcmin$)\footnotemark[1] using a combination of young and old
isochrones ($\log$ (age [yr]) = 10.1, 9.8 and 9.6, we will use these units when referring to age throughout the paper) with two different
metallicities ($[$Fe/H$]=-2$ and $-1$), allowing us to probe a more
extended parameter space than the one typically explored. We use
distance moduli between 16 and 21, which lets us explore the halo for
satellites between $\sim 16$ kpc and $\sim150$ kpc, and use an outer
kernel of $10\arcmin$, whose small size is only possible due to the
small inner kernel size. This allows the large kernel to be sensible
to rapidly changing areas like the outskirts of the Magellanic Clouds
or the galactic disk.

\footnotetext[1]{The smaller kernel size was picked based on available resources
at the time.}

\subsection{Correction for small number of counts}

In the satellite search procedure described in 
\citet{2018MNRAS.475.5085T} it is implicitly assumed,  
that the result of the convolution of density maps is Gaussian distributed.
However this approximation is bound to break if the number of stellar tracers
within the inner kernel drops low enough. This can happen if a
particularly rare type of stars is used, e.g. Blue Horizontal Branch stars
(BHBs), or, simply when the kernel size is small enough, like in the case
presented in this work. This non-Gaussianity of the distribution will 
affect the the tail probability caclulations and if ignored will lead to 
overestimated statistical significances that produce an increase in 
the number of false positive detections.

In what follows, we provide a simple correction for the significance
estimation in the case of low number of counts.  This is achieved by computing
the statistical significance values (Z-values) for an array of simulated datasets and 
comparing those to the true probabilites of a random fluctuation to find an empirical relation between
them. Our mock data are 2D pixel grids with each pixel value following the Poisson
distribution. These are then convolved with kernels of different sizes, and
the observed significance, $S_{\rm obs}$, is computed. Next, by counting the
number of pixels above a given value of $S_{\rm obs}$ and comparing it to the
total number of pixels, we can measure the real probability - which we can in
turn transform to real significance $S_{\rm true}$ - of observing a value
larger than $S_{\rm obs}$ due to random fluctuations.
Figure~\ref{fig:Significance2} shows the behavior of the true significance
$S_{\rm true}$ as a function of the measured one. As illustrated by the
Figure, the bias induced by the assumption of Gaussianity is a strong function
of the number of stars within the kernel $N_{\rm sk}=m \sigma^2_s$, where $m$
is the average number of stars per pixel and $\sigma_s$ is the width of the
inner kernel. For example, for 10 stars per kernel, while the observed
significance is reasonably high $S_{\rm obs}\sim 5$, the true significance is
much lower with $S_{\rm true}\sim 4$. The bias is exacerbated when the
number of stars per kernel plummets as low as $\sim$1: the true significance
drops to $\sim$3.

To correct for this effect, we find that the dependence of $S_{\rm true}$ as a
function of $S_{\rm obs}$ and $N_{\rm sk}$ can be empirically fitted by the
following model:

\begin{equation}\label{eq:Sig2}
S_{\rm true}=\frac{S_{\rm obs}+2a N_{\rm sk}^{-a}}{1+a N_{\rm sk}^{-a}},
\end{equation}

\noindent where $a=1/\ln 10 \approx 0.434$.

As Figure~\ref{fig:Significance2} illustrates, the above formula provides a
satisfactory approximation of the evolution of the true significance across a
wide range of the observed significance values as well as the kernel sampling
values. We apply this correction to $S$ only, before estimating $S_L$.

\subsection{Satellite Modeling}\label{sec:ODC.2}

An additional step of our search algorithm consists of a detailed study of the
probability of our candidates to be a real satellite. At this point, the
candidate's structural parameters are derived, and when color information is available, distances, ages, metallicities and
luminosities are also derived. Models are fit simultaneously to the distributions
of the candidate's stars (and the associated background) in both the celestial
coordinates and the CMD \citep[see e.g.][for a similar
approach]{Martin2008,Koposov2010}.

The probability of observing a star at a position in 4D space, $\bm{\Phi}$, spanned
by 2 spatial coordinates, a color, and a magnitude,
$\bm{\Phi}=(\bm{\Phi_s},\bm{\Phi_c})=((x,y),(\rm{col},\rm{mag}))$ is:

\begin{equation}\label{eq:Prob1}
\begin{aligned}
P(\bm{\Phi}|\gamma)= & fP_s^{obj}(\bm{\Phi_s}|\gamma_s) P_c^{obj}(\bm{\Phi_c}|\gamma_c)+\\
                     & (1-f)P_s^{bg}(\bm{\Phi_s}|\gamma_s)P_c^{bg}(\bm{\Phi_c}|\gamma_c),
\end{aligned}
\end{equation}

\noindent where $f$ is the fraction of stars that belong to the object, the
suffixes $s$ and $c$ refer to the coordinates on the sky and the CMD,
respectively. $\gamma$ is a shorthand for the model parameters, $bg$ refers to
the background model, and $obj$ refers to the satellite model.

\subsubsection{Celestial distribution model}

To avoid distortions in the shape of the models, the spatial fitting of the
candidates is performed in a gnomonic projection of the equatorial
coordinates. In the projected space $\bm{\Phi_s}=(x,y)$, the object is modeled
as a 2D elliptical plummer sphere:

\begin{equation}
  P_s^{obj}(\bm{\Phi_s}|\gamma_s)=\frac{1}{\pi a^2 \left(1-e\right)}\left(1+\frac{\tilde{r}^2}{a^2}\right)^ {-2},
\end{equation}

\noindent where $\tilde{r}^2=\tilde{x}^2+\tilde{y}^2$ and

\begin{equation}
\begin{bmatrix}
  \tilde{x} \\
  \tilde{y}
\end{bmatrix}
=
\begin{bmatrix}
  \cos{\theta}/(1-e) & \sin{\theta}/(1-e)  \\
  -\sin{\theta}      & \cos{\theta}
\end{bmatrix}
\begin{bmatrix}
  x-x_0 \\
  y-y_0
\end{bmatrix}.
\end{equation}

\noindent Then, the spatial model has 5 parameters: the center of the plummer
profile $(x_0,y_0)$, the half light radius $a$, the ellipticity $e$, and the
orientation of the ellipse $\theta$.

We model the background stellar distribution with a bilinear distribution of
the form:

\begin{equation}
  P_{s}^{bg}=\frac{1}{N_s^{bg}}(p_1x+p_2y+1),
\end{equation}

\noindent where $p_1$ and $p_2$ define the strength of the gradient in the
east-west and north-south directions respectively, and $N_b$ is defined so
$P_{s}^{bg}$ is normalized over the modeled area. In total, the spatial model
has 7 free parameters, 5 for the objects and two for the background.

\subsubsection{Color Magnitude distribution model}

When color information is available (i.e. when at least two magnitudes are
available), we model the distribution of the stars in the CMD space by binning
it into a 2D histogram and quantifying the probability of finding a star, from
background or satellite, in each bin. For the satellite, the PDF is defined
based on the mass function of the PARSEC isochrones \citep{Bressan2012}.
Specifically, the probabilities of finding a star in each bin are found in
three steps: First, we select an isochrone with a given metallicty and age,
and shift it to an specific distance. Second, we find the number of stars
along the isochrone track - as given by the isochrone mass function - and
populate the binned CMD. Finally, we convolve the populated CMD with the
corresponding photometric errors - directly drawn from the data - and
normalize the binned CMD so the integral over it is 1. By this definition, the
CMD model of the satellite has only three free parameters, namely, the
isochrone age, the isochrone metallicity, and the distance modulus.

For the background model of the CMD space, the process is much simpler. We
create it empirically from the stars around the candidate, and bin them in a
2D CMD histogram with the same shape as the one defined for the satellite
model. Then, after convolving the histogram with a small Gaussian kernel to
make it smoother, we normalize it so the integral sums to 1.

Summarizing, the full model takes 2D positions on the sky and two magnitudes
as input, and has a total of 11 free parameters. These include the isochrone
parameters - age, metallicty, and distance - that define the CMD model. The
spatial model is determined by the center, size, ellipticity, and position
angle of the plummer model, plus two extra parameters to define the
background. And finally, the fraction of stars that belong to the satellite. To fit the model to each candidate, we select the stars within one
outer kernel from its center, and find the maximum likelihood solution, first for a
background only model (i.e. $f=0$), and second for the full model described
above. This allows us to compute the log-likelihood difference, $\Delta L$ (of
the model that contains the satellite versus the model composed of background
only), giving a new significance criterion to assess if the overdensity correspond to a
real stellar system or not.



\subsection{Nine new satellites}

We applied the algorithm with the mentioned setup to the first data
release of GAIA \citep{2016A&A...595A...2G}, the first data release of
PanSTARRS \citep{2016arXiv161205560C}, the LMC/SMC bridge survey by
\citet{2018ApJ...858L..21M}, the DES year 1 catalogue produced by
\citet{Koposov2015}, ATLAS \citep{Shanks2015}, and the SDSS data
release 9~\citep{2012ApJS..203...21A}. For every survey, we then
create a list of candidates that have $S_L>4$, to ensure they are
isolated compact objects, and sort them by significance. The impact of
such selection criteria can be seen in Figure~\ref{fig:sslchap2},
which shows $S$ versus $S_L$ for all objects detected in Gaia with
$1\arcmin$ kernel. Previously known genuine stellar systems are shown
in blue, unknown detections in black. The systems presented in this
work (see below) are shown in red. The importance of $S_L$ for small
$S$ is evident. A simple cut in $S_L$ can remove most of the - alleged
- false positives without removing many good detections, allowing for
a clean and manageable list of candidates. To avoid the need of follow
up, and given that one expects to see a counterpart in the image for
the compact objects sought from this search, we only picked candidates
that were almost unequivocally confirmed by visual inspection. Then,
by going through the lists of candidates - and visually inspecting the
first hundred candidates in each survey - we were able to identify 8
unknown highly likely real stellar systems. To this list we also add a
serendipitously discovered globular cluster that lies in a chip gap in
the original DES catalogue generated by \citet{Koposov2015}.

Of the 9 objects selected, 5 correspond to discoveries made with
Gaia. As pointed out by \citet{2017MNRAS.470.2702K}, Gaia possess
unique capabilities that make it very well suited to search for
stellar systems resolved into stars. In particular, its high
resolution allows for an efficient star/galaxy separation and star
detection in crowded areas. Also, the multi-epoch observation strategy
allows for the removal of spurious detections, like the ones produced
by bright stars. The last point proved to be particularly relevant
since the brightest star in the sky was responsible for hiding a very
prominent star cluster \citep[i.e. Gaia~1][]{2017MNRAS.470.2702K} - in
a frequently visited area of the sky - until now. One of the clusters
presented here, Gaia~7, is also obscured by a bright star.

%


\subsection{Multi-band photometry analysis}

The availabilty of multi-band photometry allows for the inference of physical and stellar population properties of the stellar systems. Here we present the result of the modelling of the nine star cluster candidates using the isochrone fitting technique. This fit allows us to
estimate satellite's distances and luminosities. For the five systems initially found in Gaia data, we used deeper available data. Specifically, three of the Gaia satellites
had Pan-STARRS data, the other two fell into the footprints of SMASH, in the case of Gaia~3, and DeCAPS, in the case of Gaia~6. Of
the remaining satellites, 2 were discovered in DES, one was discovered
in the MC bridge survey \citep{2018ApJ...858L..21M}, and the last one was
detected in the Pan-STARRS data. The detailed
summary together with the main physical properties inferred can be
found in Table \ref{tab:propwithiso}.


{\bf To~1} is the most prominent of the discoveries presented in this
paper. It is identified in the MC bridge survey and boasts an unambiguous
detection significance of 16.5. The discovery plot for To~1 is given
in Figure \ref{fig:To1}. Top panels display the spatial information,
and the bottom panels show the CMD information. To~1's signal is
obvious across all panels, with a CMD harboring a strong main sequence
turn-off (MSTO) together with a small number of plausible RGB
candidates. The best-fit isochrone, shown in red, is consistent with
an old and metal poor population at $\sim43.5$ kpc, placing To~1 only
$\sim7$ kpc away from the LMC. It has an absolute magnitude of
$M_V\sim -1.6$ and a physical size of $\sim3.5$ pc, thus placing it in
a region of the size-luminosity plane below the main locus of globular
clusters. This region have been gradually populated with various
discoveries in the last 10 years
\citep[e.g.][]{Koposov2007,Balbinot2013,2016ApJ...820..119K}. Its position in the
sky, as well as its physical properties, suggest that To~1 may be
associated with the MCs.

{\bf Gaia~3}'s discovery is shown in Figure~\ref{fig:Gaia3}. Originally found in Gaia with a significance in excess of 8, Gaia~3 is the second most prominent detection presented in this work. Coincidentally, Gaia~3 happens to lie in one of the pointings of the SMASH survey \citep{2017AJ....154..199N}, but the associated data is not available as part of their first data release. Nevertheless, the stacked images are publicly available through the NOAO Science Archive (NSA). Although data for this part of the sky is already provided in the NOAO source catalogue \citep[NSC,][]{2018AJ....156..131N}, most sources in the central parts of Gaia~3 were missing, probably due to crowdedness. Consequently, we re-analyzed the SMASH g-band and r-band stacked images with a simple SExtractor+PSFex combo \citep{Bertin1996}, allowing for higher de-blending, and calibrated it using the NSC. With the resulting catalogue, we can easily see Gaia~3 in the density maps and in the CMD. Indeed, Gaia~3 is well fit by our stellar+spatial model (with a likelihood difference of $\Delta L\sim 90$). The best fit reveals that Gaia~3 is a compact $r_h\sim0.4\arcmin$ relatively luminous stellar system ($M_V\sim-3.3$) located at $D_H\sim49$\,kpc with $\log(age) \sim 9.1$ and a metallicity of $[$Fe/H$]\sim-1.8$. Located $\sim 6$ degrees away from the LMC, its position and
characteristics reflects those of the LMC cluster system, suggesting a
Magellanic origin. This cluster is perhaps a prime example
of how the combination of Gaia, with its superb astrometry, plus the
small spatial scales used in this work can reveal previously
undiscovered - but very prominent - stellar associations.


\begin{figure*}
    \includegraphics[width=\textwidth]{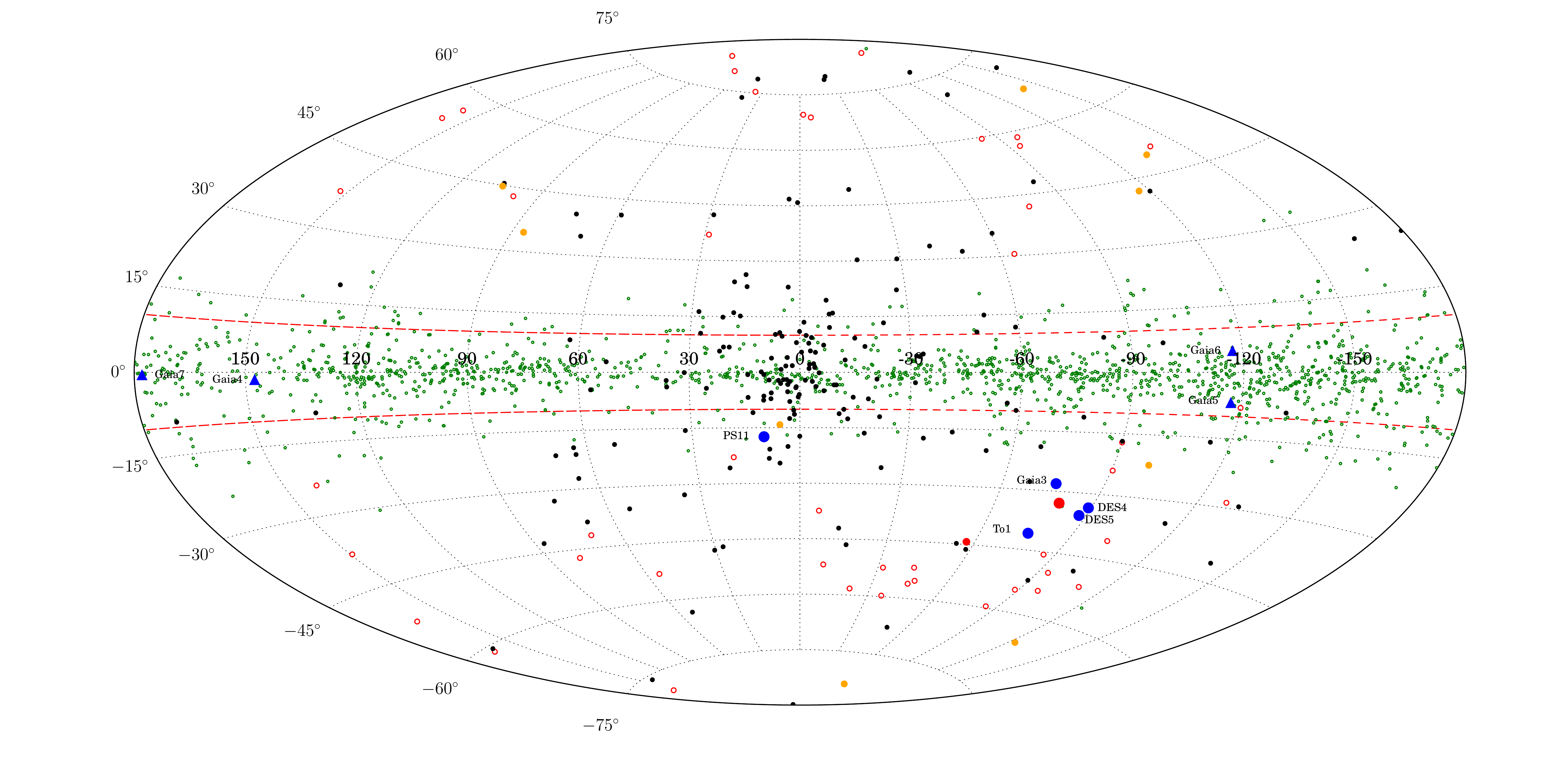}
    \caption{Spatial distribution of GCs and dwarf galaxies around the MW in galactocentric coordinates. Yellow markers shows the position of the Classical dwarfs, open red circles shows the position of other MW galaxies (from \citet{McConnachie2012} plus Crater~2
      \citep{Torrealba2016}, Aquarius~2 \citep{2016MNRAS.463..712T},
      DESJ0225+0304 \citep{2017MNRAS.468...97L}, Pictor~II
      \citep{2016ApJ...833L...5D}, Virgo~I \citep{2016ApJ...832...21H},
      Cetus~III \citep{2018PASJ...70S..18H}, and Car~II and Car~III \citep{2018MNRAS.475.5085T}), black dots the position of MW GCs \citep{Harris2010,Belokurov2010,Munoz2012,Balbinot2013,Kim2015a,Kim2015b,2016ApJ...820..119K,Laevens2015,Weisz2016,Luque2015,2017MNRAS.468...97L,2016ApJ...830L..10M,2017MNRAS.470.2702K}, and green open dots show the position of MW OCs \citep[from][]{2013A&A...558A..53K}. The new discoveries are shown in blue. The red dashed lines shows $|b|=10\deg$. 4 of the new discoveries, namely, Gaia~3, DES~4 and 5, and To~1 are very close to the LMC, and likely members of its GC population. The remaining 4 systems, Gaia~4 to Gaia~7, are located well within the disk, and are likely open clusters.}
    \label{fig:spacedist}
\end{figure*}

{\bf Gaia~4} is shown in Figure~\ref{fig:Gaia9}. Like Gaia~7, it is
located very close to the Galactic disk ($b\sim-1.5$) in a region of
high extinction and within the Pan-STARRS footprint. The overdensity
of stars can be easily seen in the Pan-STARRS density maps, and the
photometry reveals a well-defined MS. As a result we are able to get a
good fit of its structural and stellar population properties. Gaia~4 is
found to be composed of a compact $(r_h\sim1.17\arcmin)$ stellar
population at $DM\sim13.7$ with $\log(age) \sim 9$ and a metallicity
of $[$Fe/H$]=-0.1$, and it is the second most luminous object
presented in this work with $M_V\sim-2.4$. But even at this
luminosity, and a relatively high significance ($S\sim7.5$) in our
Gaia searches, it was still missing in previous Pan-STARRS
searches. Perhaps this is due to a combination of the satellite's
small size and a high (and variable) extinction in the region. Given
its position in the disk, low luminosity, small size, and the lack of
a well defined center in the false-color image, it looks like Gaia~4
is probably another Galactic open cluster.

\begin{table*}
    \caption{Properties of clusters with isochrone fit}
    \label{tab:propwithiso}
    \centering
    \begin{tabular}{lrrrrrrrrrr}
        \hline
        Name   & RA        & Dec       & S    & $\Delta L$ & $r_h$      & DM   & $D_h$& $M_V$& $r_h$&$N_{\rm{star}}\,\,^a$  \\
               & (deg)     & (deg)     &      &            &($\arcmin$) &(mag) &(kpc) & (mag)& (pc) &      \\
        \hline
        To~1    & 56.08255  & -69.42255 & 16.5 & 88.9       & 0.27       & 18.2 & 43.6 & -1.6 & 3.45 & 33   \\
        Gaia~3  & 95.05864  & -73.41445 & 8.2  & 89.8       & 0.53       & 18.4 & 48.4 & -3.3 & 7.45 & 86   \\
        Gaia~4  & 56.36793  & 52.89297  & 7.5  & 46.1       & 1.17       & 13.7 & 5.4  & -2.4 & 1.85 & 58   \\
        PS1~1   & 289.17121 & -27.82721 & 6.3  & 49.6       & 0.55       & 17.4 & 29.6 & -1.9 & 4.69 & 42   \\
        Gaia~5 & 110.79779 & -29.71947 & 6.2  & 26.7       & 1.01       & 14.2 & 6.8  & -0.1 & 2.01 & 20   \\
        Gaia~6 & 122.09798 & -23.70648 & 6.1  & 55.4       & 1.20       & 11.9 & 2.4  & 0.2  & 0.85 & 101  \\
        Gaia~7  & 84.69075  & 30.49822  & 5.5  & 15.5       & 0.70       & 13.0 & 4.0  & -1.8 & 0.82 & 13   \\
        DES~4   & 82.09501  & -61.72369 & 5.1  & 13.0       & 0.83       & 17.5 & 31.3 & -1.1 & 7.58 & 42   \\
        DES~5   & 77.50351  & -62.58046 & -    & 13.4       & 0.18       & 17.0 & 24.8 & 0.3  & 1.31 & 10   \\
        \hline
        \multicolumn{11}{l}{$^a\,\,N_{\rm{star}}$ is the measured number of the system member stars above the limiting magnitude}\\
    \end{tabular}
    \caption{test}
\end{table*}


{\bf PS1~1/Prestgard~64} is the only detection procured when applying the search
algorithm to the Pan-STARRS data.PS1~1 was also independently identified by amateur astronomer Trygve Prestgard in DSS data in 2016 as Prestgard~64\footnotemark. In PS1 data, it can be confidently reported as a
genuine stellar system without the need of a follow-up as shown in
Figure~\ref{fig:PS11}. At the center of PS1~1 an evident stellar
overdensity is visible both in the cutout images as well as the
density maps. The main feature of PS1~1 in the CMD diagram is a clump
of stars at $r\sim21$. The maximum likelihood fitting favors a model in which this feature is explained by the MSTO of a stellar population at $DM=17.4$, which yields a luminosity of
$M_V\sim-1.9$ and a half light radius of $r_h\sim4.7$ pc, which is consistent with both a GC or an OC classification.

\footnotetext{see also: https://skyhuntblog.wordpress.com/my-asterisms-and-possible-star-clusters/}

{\bf Gaia~5 and 6} discovery plots are shown in Figure \ref{fig:Gaia10} and
\ref{fig:Gaia11}. The two satellites fall
within the Pan-STARRS footprint, but were missed by previous searches
- and our own Pan-STARRS search - likely because of their proximity
to the Galactic disk ($|b|\lesssim5\deg$). Gaia~6 is also within the DEcam Plane Survey footprint \citep[DECaPS][]{2018ApJS..234...39S}, and we use their first public data release to measure the physical and stellar population properties of Gaia~6. Both Gaia~5 and ~6 share very similar
properties: they are located close to the sun ($D_h\sim5$ kpc), are
well embedded into the Galactic disk, and have sizes of the order of 1
pc, and absolute magnitudes hovering just above $M_V\sim0$. Based on
their location, structural properties, and the appearance of the image
cutouts, we can confidently classify them as Galactic open
clusters. We note that a classical Cepheid is sitting close to the
center of Gaia~5. The star has a period of 3.33 days and a an average
magnitude of $V\sim12.95$ \citep{2006OEJV...54....1O}. Using the
period-luminosity relation for classical Cepheids
\citep{1997MNRAS.286L...1F}, we find $DM\sim15.85$ for the star. This
is $\sim1.7$  magnitudes off from the the DM found for the best fit
isochrone, hence, we conclude that the Cepheid and the open
cluster are likely unrelated.


{\bf Gaia~7} discovery is summarized in Figure~\ref{fig:Gaia8}. Similar to Gaia~1
\citep[][]{2017MNRAS.470.2702K}, the very first MW satellite
discovered using the Gaia data, Gaia~7 is located next to a
bright star. Regions next to very bright objects are severely affected
by artifacts - due to saturation and diffraction - in other
photometric surveys, but thanks to Gaia, which excels at the removal
of such spurious detections \citep[see e.g.][]{2016A&A...595A...3F},
these regions are now accessible. Gaia~7 is located within the
Pan-STARRS footprint, but while it is visible in its density maps, it was missed due to the proximity to the very bright star. Nevertheless, it is still clearly visible in the Pan-STARRS data. Indeed, as shown in Figure~\ref{fig:Gaia8}, Gaia~7 is easily visible both in Pan-STARRS images and density maps. Gaia~7 also features a well defined main sequence, to which we fit an isochrone while simultaneously model its on-sky distribution. As a result, we find that Gaia~7 is consistent with a relatively young population (log age$\sim 7.85$) with $[$Fe/H$]=0.1$ at $DM\sim 13.1$. We measure its half light radius to be $r_h\sim0.42$pc and its absolute magnitude $M_V\sim2.8$. These properties and its location let us confidently classify it as a Galactic open cluster. However, please note that the proximity of the bright star, as well as the high extinction in the region - E(B-V)$\sim 1.2$, which means almost 4 magnitudes of extinction in $g$ - could induce strong systematics in either the estimation of physical as well as stellar population properties.

{\bf DES~4 and DES~5}, the two satellites discovered in the DES data,
are both only a few degrees away from the LMC. Their discovery plots
are shown in Figure~\ref{fig:DES3} and Figure~\ref{fig:DES4}
respectively. Their CMDs are dominated by the LMC stars, which makes
the isochrone mask of the target population very similar to the
background. For this reason, DES~4 significance is quite low, and
DES~5 was not even found in the search. However, both overdensities
are extremely compact, and feature a conspicuous stellar overdensity
in the false color images. In fact, DES~5 was serendipitously found
while looking at images around the LMC to study the performance of the
search in this region. Bear in mind that DES~5 was located in a chip
gap in the original \citet{Koposov2015} catalogue, and hence missed
from the searches, but with the first data release from DES
\citep[DR1,][]{2018ApJS..239...18A}, the object is now completely
sampled. In both cases the Hess difference diagram reveals an excess
of stars that appears to be a part of a stellar population younger
compared to the bulk of the LMC population. These CMD features are
captured by the satellite model which we found to be more likely than
the background-only model by $\Delta L\sim13$. DES~4 has $M_V\sim-1.1$
and $r_h\sim7.58$ pc, placing it at the interface between GCs, OCs,
and Ultra-Faint Dwarfs (UFDs) in the size luminosity diagram. DES~5
fit is more compact ($r_h\sim1.31$ pc) and much fainter
($M_V\sim0.3$). Note that in both cases the high concentrations of
stars in the object is likely affecting the photometry. This
effectively means that some of the member stars missing from the
catalogues. As a result, it is possible that both objects have their
absolute luminosities underestimated by few $10^{-1}$ mags. For example, if
DES~4 had $30\%$ more stars, its luminosity will go up from $M_V=-1.1$
to $M_V=-1.4$.

\begin{figure*}
    \includegraphics[width=\textwidth]{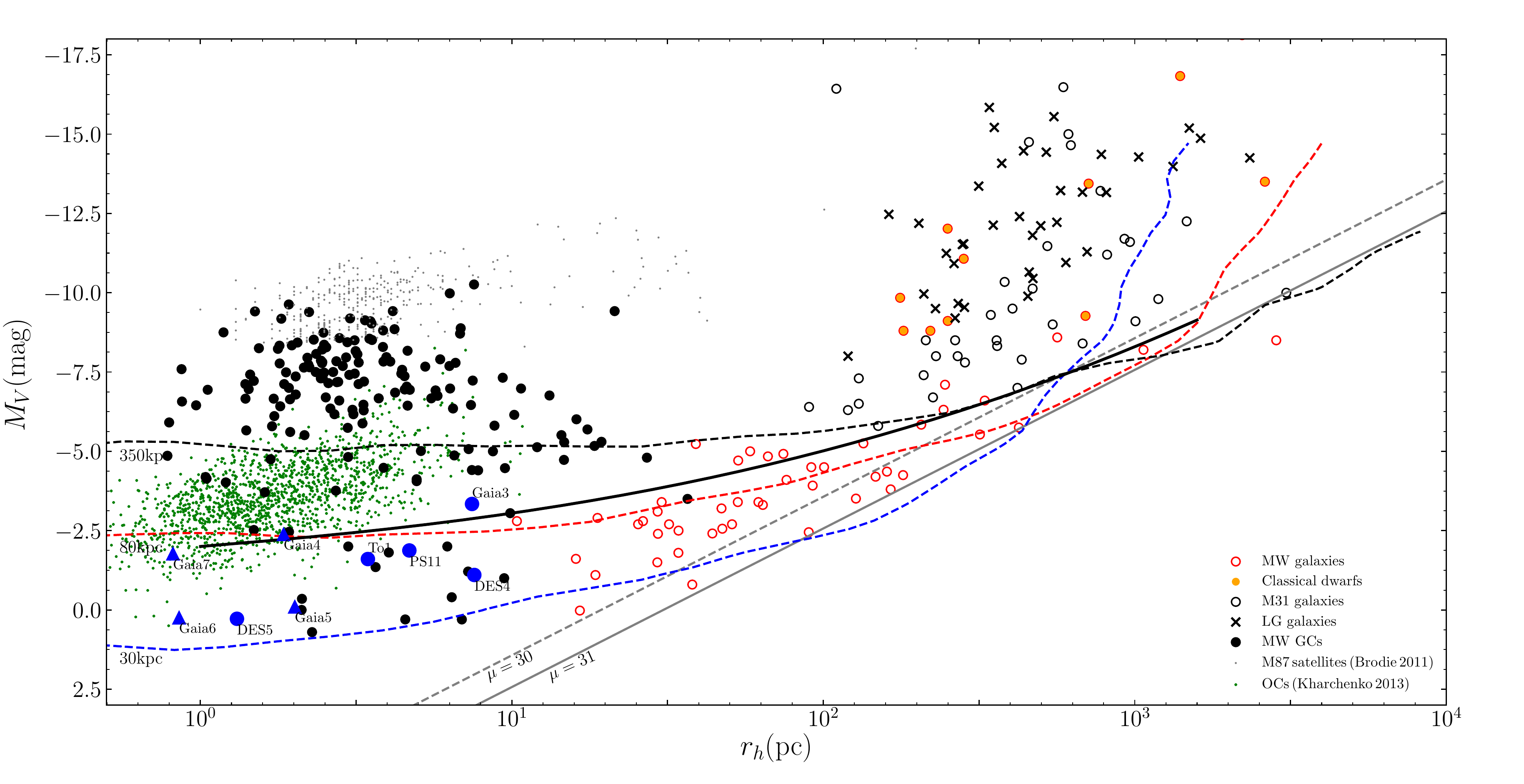}
    \caption{Absolute versus half light radius diagram. The markers are the same as Figure~\ref{fig:spacedist}, plus M31 galaxies shown with black open circles, LG galaxies shown as black crosses \citep[both from][]{McConnachie2012}, M87 satellites shown in gray dots \citep[from][]{Brodie2011}, and OCs are shown as open green dots \citep[from][]{2013A&A...558A..53K}. Lines of constant surface brightness at $\mu=31$ and 30 are shown in gray continuous and dashed lines respectively. These roughly represent the detection limits of current surveys \citep{Koposov2008,Koposov2015}. Lines of effective detection limits of the algorithm presented in this paper are shown as dashed blue, red, and black lines, representing the limits for satellites located at distances of 30, 80, and 350 kpc, respectively. The black line shows an effective limit of known satellites detected in Gaia DR1, above the line most satellites are detected, and below, most are missed. The new discoveries are shown in blue, with the objects sitting in the Galactic disk marked as triangles. Interestingly, they are mostly located in a sparsely populated region of the diagram, where classification is uncertain. The new discoveries are part of the population of the smallest and lowest luminosity stellar systems in the MW halo.}
    \label{fig:mvrhclusters}
\end{figure*}

\section{Conclusions}

We have presented a systematic search for compact stellar systems
across different photometric surveys. Building on previous efforts
\citep[specially][]{Koposov2008}, we have added features to improve
the detection of stellar systems, particularly at the faint
end. Specifically, we provide a recipe to deal with the small number
statistics, which allows for a search of very small objects without
flooding the candidates list with false positives.

We have performed a search for satellite systems with sizes smaller
than $1\arcmin$ in most available wide surveys, including Gaia,
Pan-STARRS and DES. Even though these datasets have been mined for
satellites before, we were able to detect 9 new stellar systems,
highlighting the importance of the improvements in the detection
procedure and the extension of the probed parameter space.
Figure~\ref{fig:spacedist} shows the distribution of the new
discoveries on the sky. New satellites are marked in blue. About half of the new satellites are close to the
LMC, and are likely part of its star cluster system.  The other half are located in the Galactic disk, where the bulk of the 
population of open clusters is sitting, hence likely OCs.

As Figure~\ref{fig:mvrhclusters} demonstrates, the newly discovered
satellites can all be found in the sparsely populated region in the
left corner of the size-luminosity space.  The new objects located in
the Galactic disk, likely open clusters, are shown with
triangles. Open clusters luminosities and sizes, estimated from the
\citet{2013A&A...558A..53K} open cluster catalogue, are shown in
green, and forms a diagonal cloud in the diagram. This distribution
closely resembles the two most characteristics limits in catalogue
based searches of stellar systems, namely the limiting in magnitude
and the limiting in surface brightness. This suggests that the current
list of OCs is displaying strong selection effects. In this diagram,
OCs are sitting just below GCs, but while they are both fainter and
smaller, it appears that there is a smooth transition between the two
groups, instead of a clear distinction. The new discoveries discussed
in this paper are the faintest population in this diagram, sitting
just below the OC loci, suggesting that they are perhaps just an
extension of the OC population.  Interestingly, both the OC catalogue
and our searches are extremely incomplete in this region, especially
in distance: none of the objects presented in this work are further
than 50 kpc, and the OC catalogue is only complete out to $\sim2$
kpc. It is then possible, if not expected, that a large population of
faint but extended OCs is waiting to be found. Older OCs appear to be
fainter and larger in the \citet{2013A&A...558A..53K} catalogue, both
traits that make them more difficult to find. This is in line with the
apparent shortage of old OCs in the MW
\citep{2014A&A...568A..51S}. However, given the results of our search,
the objects similar to those displayed here are common in the MW, thus
starting to account for the alleged shortage.

Undoubtedly, one of the clear protagonists of this satellite search is
Gaia. By covering the whole sky, Gaia allows for a holistic view of
our Galaxy, including the bulge, the disks and the halo. Surprisingly,
not only one can detect compact objects in Gaia, but it is also
possible to detect some of the more luminous UFDs. This is illustrated
with the black line in Figure~\ref{fig:mvrhclusters}, which gives the
effective detection boundary for known MW satellites in Gaia: above
the line we can detect known satellites across a wide range of sizes,
but below the line most will be missed. While our current search in
Gaia still suffers from a lot of drawbacks, namely, e.g. an unknown
amount of spurious detections in the Gaia data (see e.g. the all sky
view from Gaia DR1), it is clear that Gaia's is a wonderful dataset to
study the halo, and particularly useful to build a homogeneous census
of stellar systems across all latitudes. It is important to note that
this search was carried out with the first data release of Gaia. The
second data release provides significant improvements to the satellite
detection capabilities. Particularly, the five-parameter astrometric
solutions for 1.3 billion sources down to $G\sim21$, as well as the
photometry in blue and red passbands for a similar number of stars are
ideal to survey the whole sky for nearby stellar systems. This can
also be augmented by the information on 500 thousand variable stars,
which is a huge step forward compared to the Gaia DR1. As we have
shown in this work, there is still much to be found in MW disk and
halo, therefore all-sky, homogeneous surveys like Gaia can make an
appreciable impact in the quest of characterizing MW structure and
dynamics. Finally, while we have been able to update the census of
stellar systems inside and in the outskirts of the MW and show that
there is still room for unknown - relatively close - systems, our
search is not completely automated.  Indeed, in this instance we are
still cherry-picking by selecting only the objects that can be
identified in the image cutouts. There is plenty of room to improve
the satellite search.

\section*{Acknowledgements}

GT acknowledge support from the Ministry of Science and Technology
grant MOST 105-2112-M-001-028-MY3, and a Career Development Award (to
YTL) from Academia Sinica. VB thanks the Cambridge Streams Club for
stimulating discussions. The research leading to these results has
received funding from the European Research Council under the European
Union's Seventh Framework Programme (FP/2007-2013) / ERC Grant
Agreement n. 308024.




\bibliographystyle{mnras}
\bibliography{biblio} 



\appendix

\section{Stellar sample selection}

Here we present the catalog-level selections applied to the different
surveys to produce clean stellar samples that were then used in the
satellite search algorithm.

For {\bf Gaia}, we applied a cut on $G$ between 7 and 21 magnitudes,
and selected stars by following \citet{2017MNRAS.470.2702K}, namely
picked objects with
$\log_{10}\left(\mathrm{astrometric\_excess\_noise}\right)<0.15(G-15)+0.25$. We
also corrected for extinction using \citet{SFD} maps and the
\citet{2017MNRAS.466.4711B} extinction coefficient. For {\bf
  PanSTARRS}, star/galaxy separation was achieved by selected objects
with the difference between PSF and KRON magnitudes smaller than 0.05
mags \citep[see][for details]{2016arXiv161205560C}. Additionally, only
objects with g and r magnitudes between 14 and 23.2 mags were
selected. We also corrected for extinction using the \citet{SFD} maps
and \citet{Schlafly2011} coefficients. For {\bf DES}, we used the
star-galaxy separation cut suggested by \citet{Koposov2015}, and
selected objects with magnitudes between 16 and 23.5
magnitudes in $g$ and $r$ filters. Extinction correction was done as above, using an
appropriate \citet{Schlafly2011} coefficient for the DES filter
set. Finally, for the data from the \citet{2018ApJ...858L..21M}
survey, we picked stars with magnitudes between 16 and 23, and used
the same star/galaxy separation and extinction correction methods as
applied to the DES data.

\section{Discovery plots for 8 star clusters}

The full list of plots summarizing the discovery of the presented
objects is given in this appendix.

\begin{figure*}
    \includegraphics[width=\textwidth]{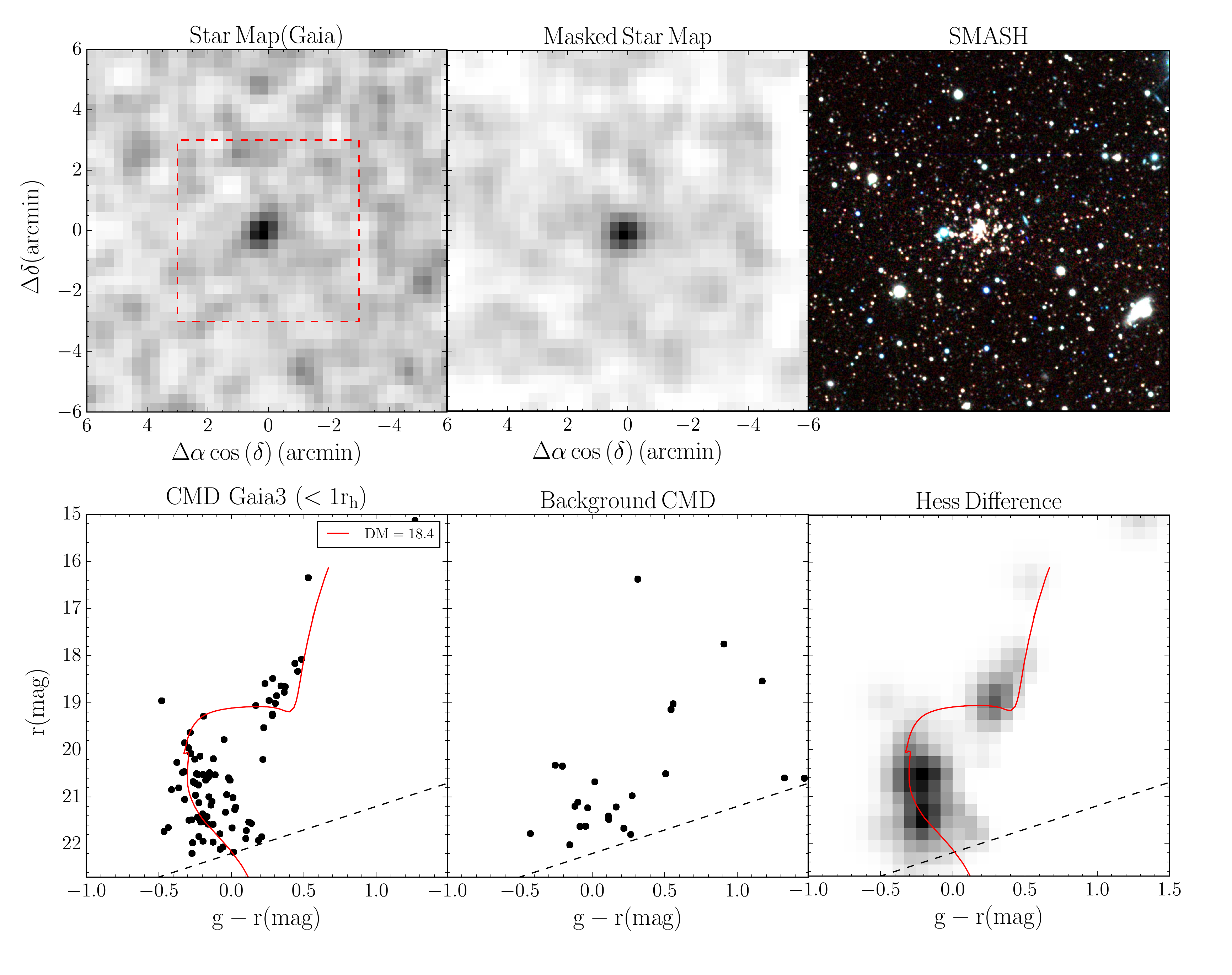}
    \caption{Discovery plot for Gaia~3. The panels are as described in Figure~\ref{fig:To1}, but with the false color image computed from the SMASH survey. Gaia~3 is only $5.8\deg$ from the LMC, and at a similar distance ($D_h\sim48$ kpc) so it is likely a GC that avoided detection due to the high concentration of stars in its center. Nevertheless, by re-analyzing the images from the SMASH survey, we were able to fit an stellar population to the object which is metal poor ($[$Fe/H$]\sim-1.8$) with an age of log age$\sim 9.1$ with a total luminosity of $M_V\sim-3.3$ and a size of $r_h\sim7.5$pc. Its properties, plus the tight concentration of stars at its center, makes Gaia~3 likely a GC associated with the LMC.}
    \label{fig:Gaia3}
\end{figure*}
\begin{figure*}
    \includegraphics[width=\textwidth]{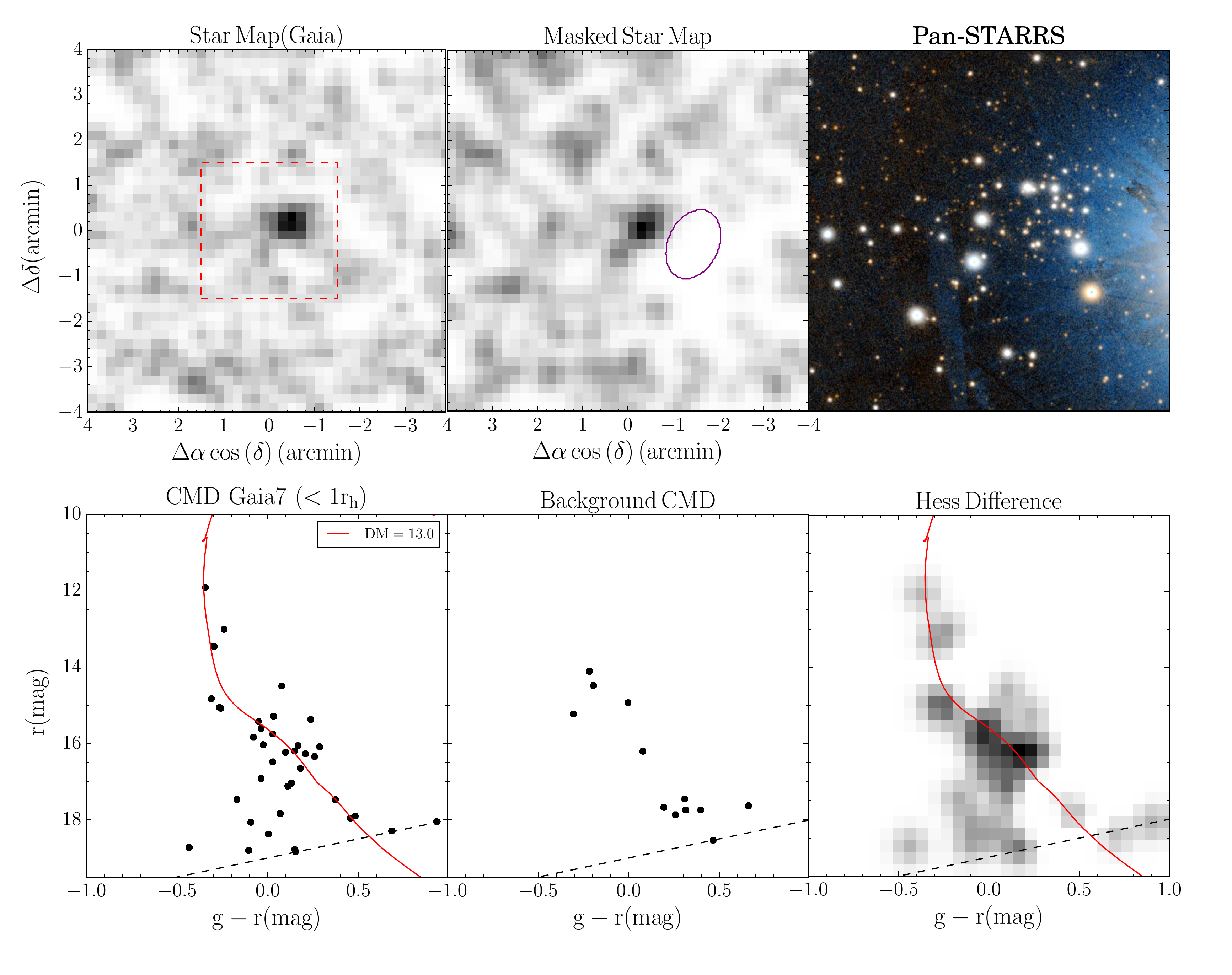}
    \caption{Discovery plot for Gaia~7. The panels are as described in Figure~\ref{fig:To1}, but with the false color image from the Pan-STARRS image server. Gaia~7 is a cluster that hides behind a bright star. In this case, the cluster is within the Pan-STARRS footprint, making it possible to fit an stellar population model to its CMD, which is fully consistent with an OC classification. The purple circle in the middle top panel marks the region around the bright star that we masked out when doing the fitting.}
    \label{fig:Gaia8}
\end{figure*}
\begin{figure*}
    \includegraphics[width=\textwidth]{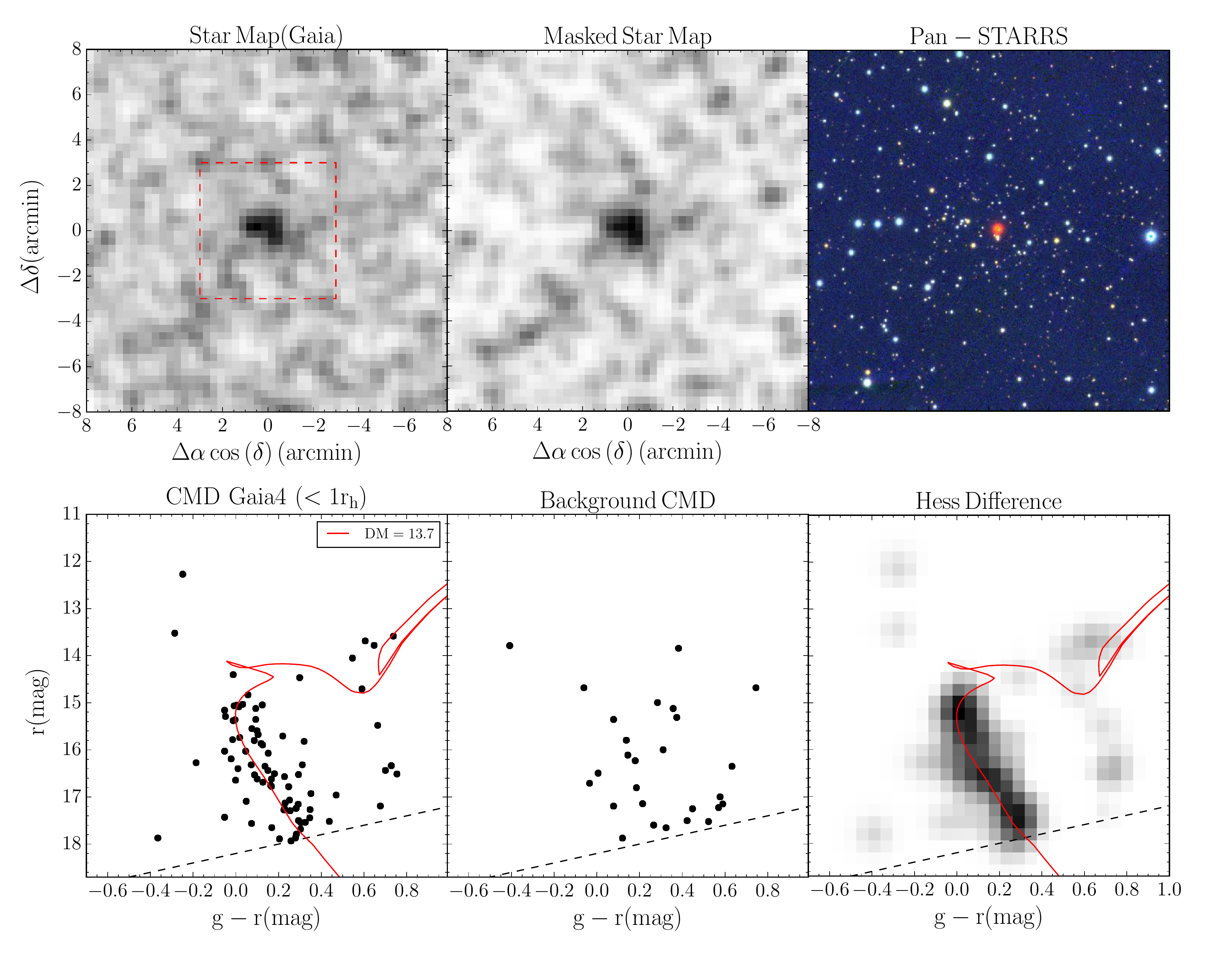}
    \caption{Discovery plot for Gaia~4. The panels are as described in Figure~\ref{fig:To1}, but with the false color image from the Pan-STARRS image server. Gaia~4 is also within the Pan-STARRS footprint, sitting in the Galactic disk ($b\sim-1.5$) at a region of very high extinction. The overdensity of stars is very prominent both in Gaia and Pan-STARRS, with a very well defined MS, making it the brightest of the objects presented in this paper with $M_V\sim-2.4$.}
    \label{fig:Gaia9}
\end{figure*}
\begin{figure*}
    \includegraphics[width=\textwidth]{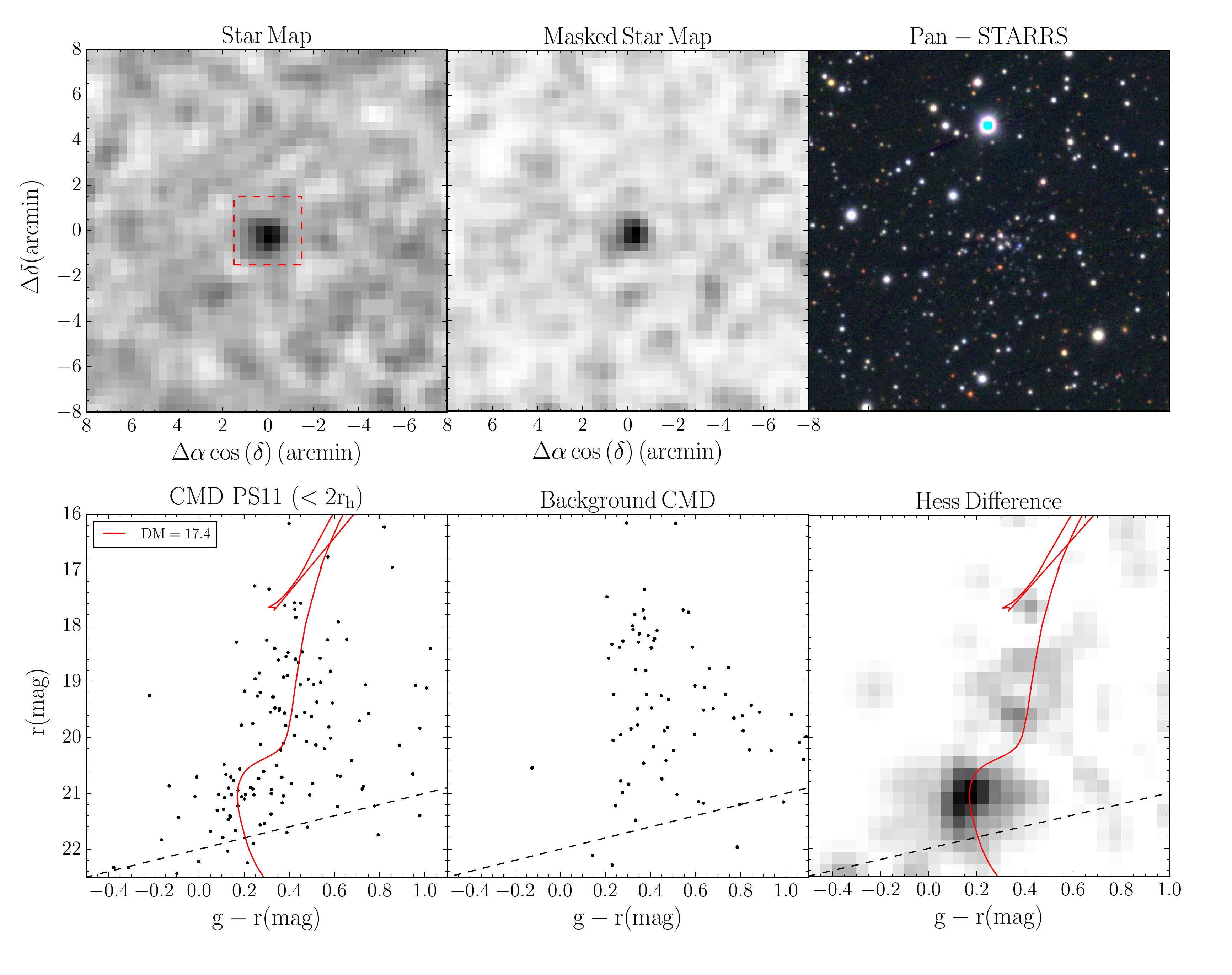}
    \caption{Discovery plot of PS1~1. The panels shows Pan-STARRS data release 1 observations, and are as described in Figure~\ref{fig:To1}. In this case, the Masked star map is only showing MSTO stars. PS1~1 spatial overdensity is evident both in the images and in the density maps. A conspicuous accumulation of stars in the CMD at the position of the MSTO, where no much background stars are expected, can also be seen.}
    \label{fig:PS11}
\end{figure*}
%


%
\begin{figure*}
    \includegraphics[width=\textwidth]{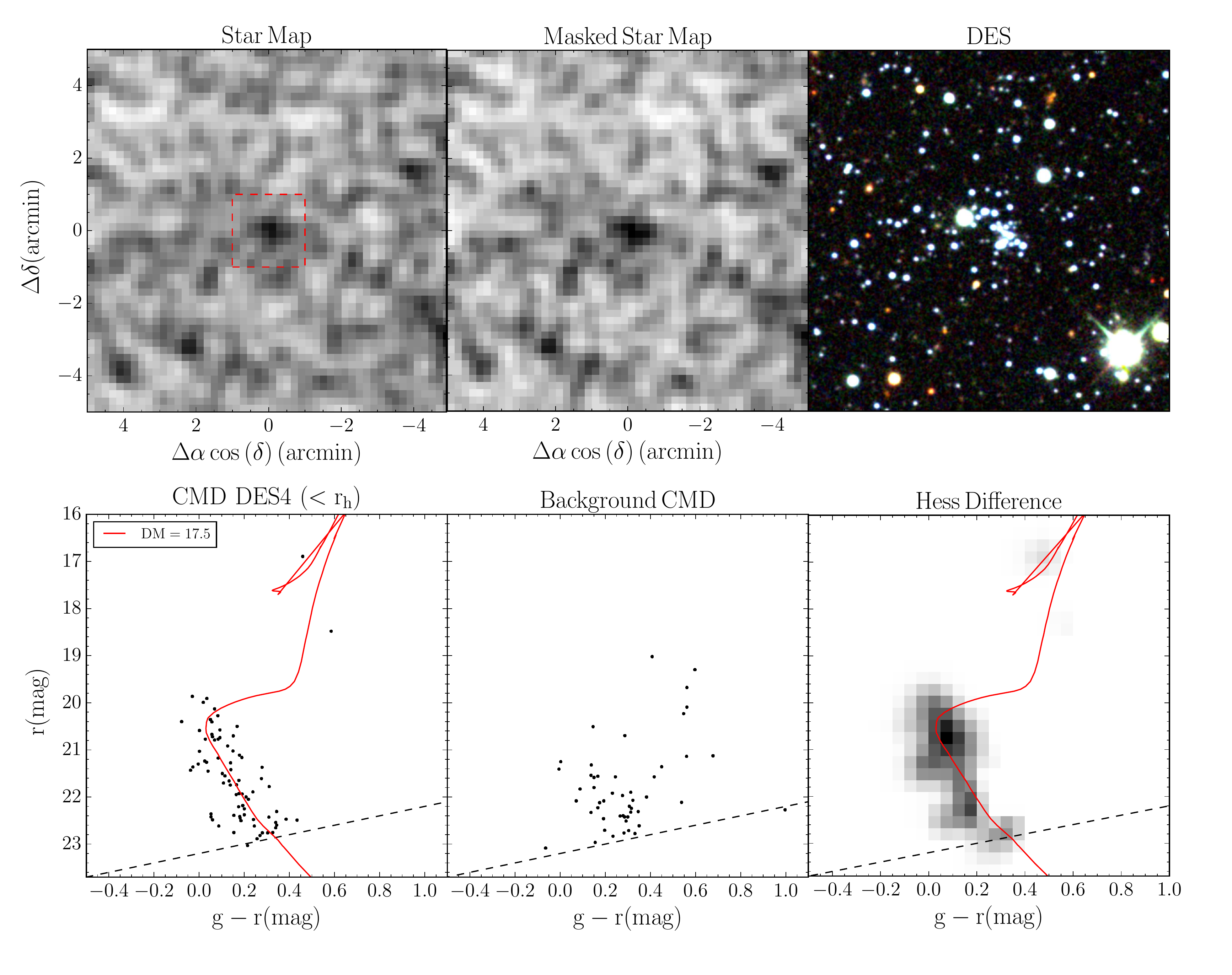}
    \caption{Discovery plot of DES~4. The panels shows \citet{Koposov2015} version of DES, and are as described in Figure~\ref{fig:To1}. Event though the density map marginally shows an overdensity, the false color image highlights a conspicuous overdensity of stars. The contrast is likely due to missing stars because of crowding. DES~4 is located next to the LMC. This can be seen in the CMDs of the object and the background. Even so, an extended MSTO reveals DES~4 in the CMDs as formed by a young population, like many of the LMC clusters. }
    \label{fig:DES3}
\end{figure*}
\begin{figure*}
    \includegraphics[width=\textwidth]{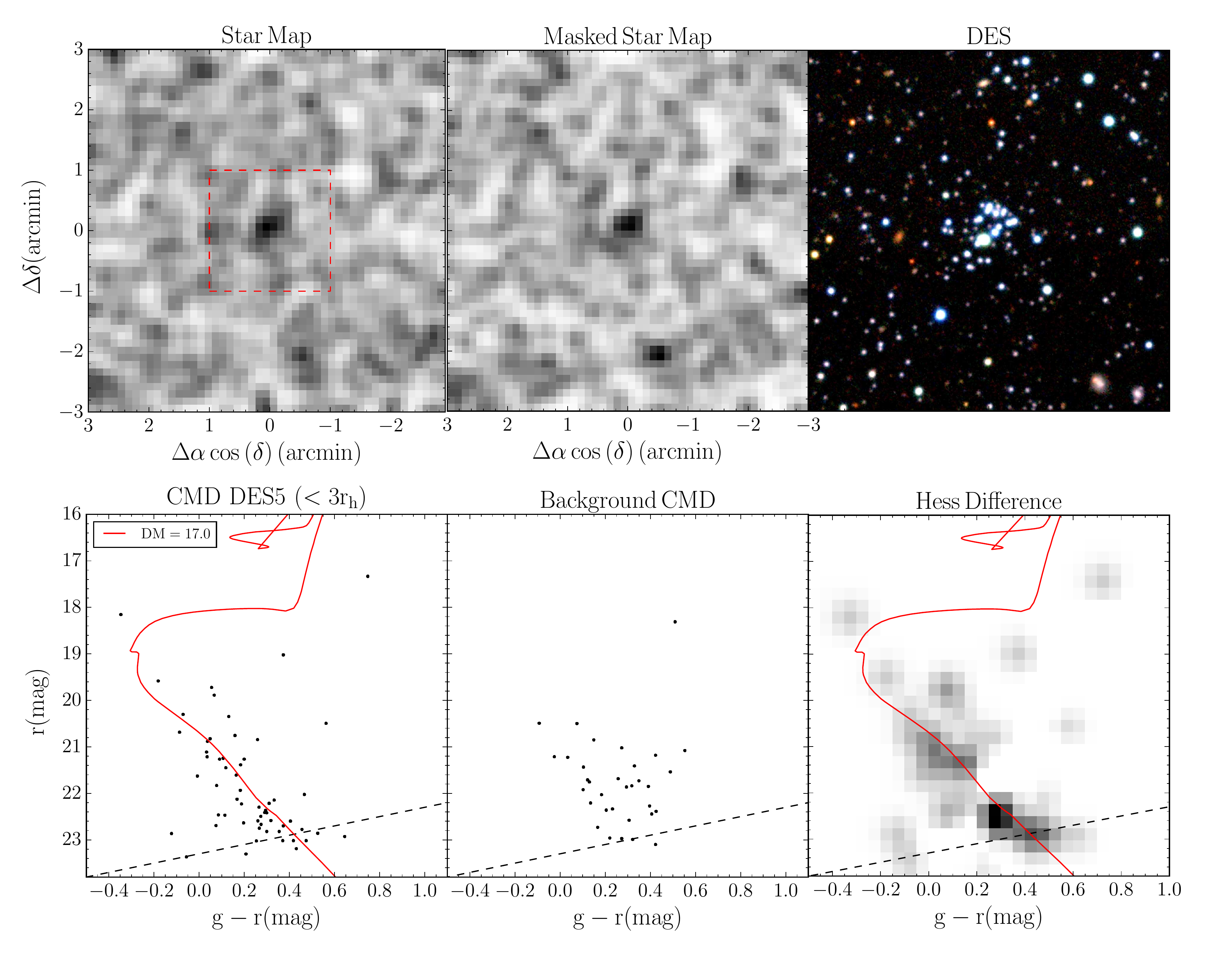}
    \caption{Discovery plot of DES~5 as seen with DES DR1 data. All panels are described as in Figure~\ref{fig:To1}. The false color image alone can confirm Des~5 as a distinct group of stars, but it is also seen as a conspicuous overdensity of stars in the density maps. The extended main sequence, compared to the background, allows to constrain the distance to Des~5 at 25 kpc. Located only $\sim 7\deg$ away from the LMC, the position of DES~5 is consistent with being part of the LMC GC population. }
    \label{fig:DES4}
\end{figure*}
\begin{figure*}
    \includegraphics[width=\textwidth]{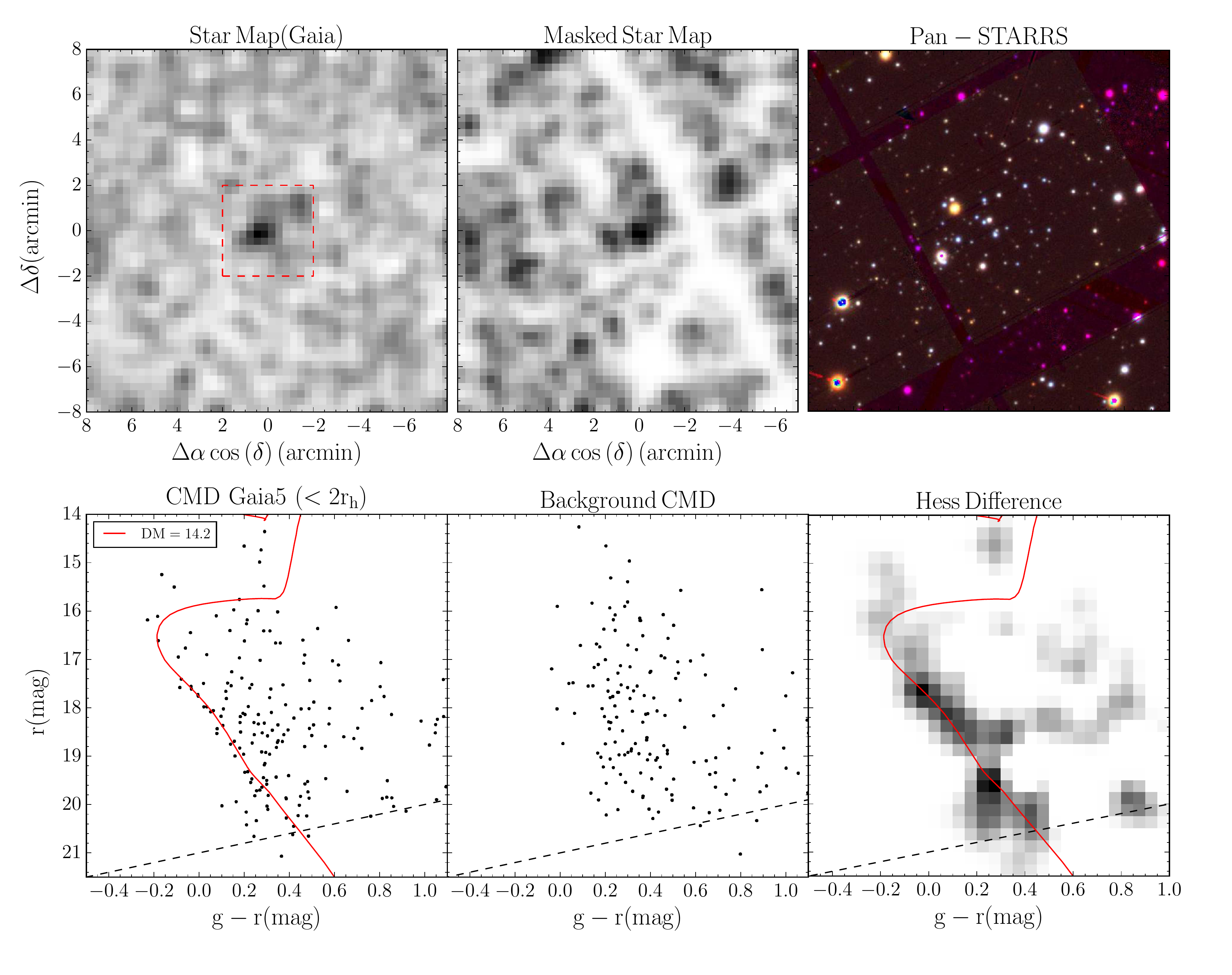}
    \caption{Discovery plot of Gaia~5. Top left panel shows a density map of Gaia stars, while the rest of the panels shows Pan-STARRS data. All panels are described as in Figure~\ref{fig:To1}. Gaia~5 reveals itself unambiguously in the Gaia density map of stars, but due to incomplete data and high extinction in the region, it was not originally found in Pan-STARRS. The hints of a distinct group of stars in the false color image, as well as the young main sequence seen in the CMDs suggests that Gaia~5 is likely a genuine satellite. Gaia~5 is located very close to the Galactic disk at $b\sim-7\deg$, added to its properties, it looks like it belongs to the Galactic disk open cluster population.}
    \label{fig:Gaia10}
\end{figure*}
\begin{figure*}
    \includegraphics[width=\textwidth]{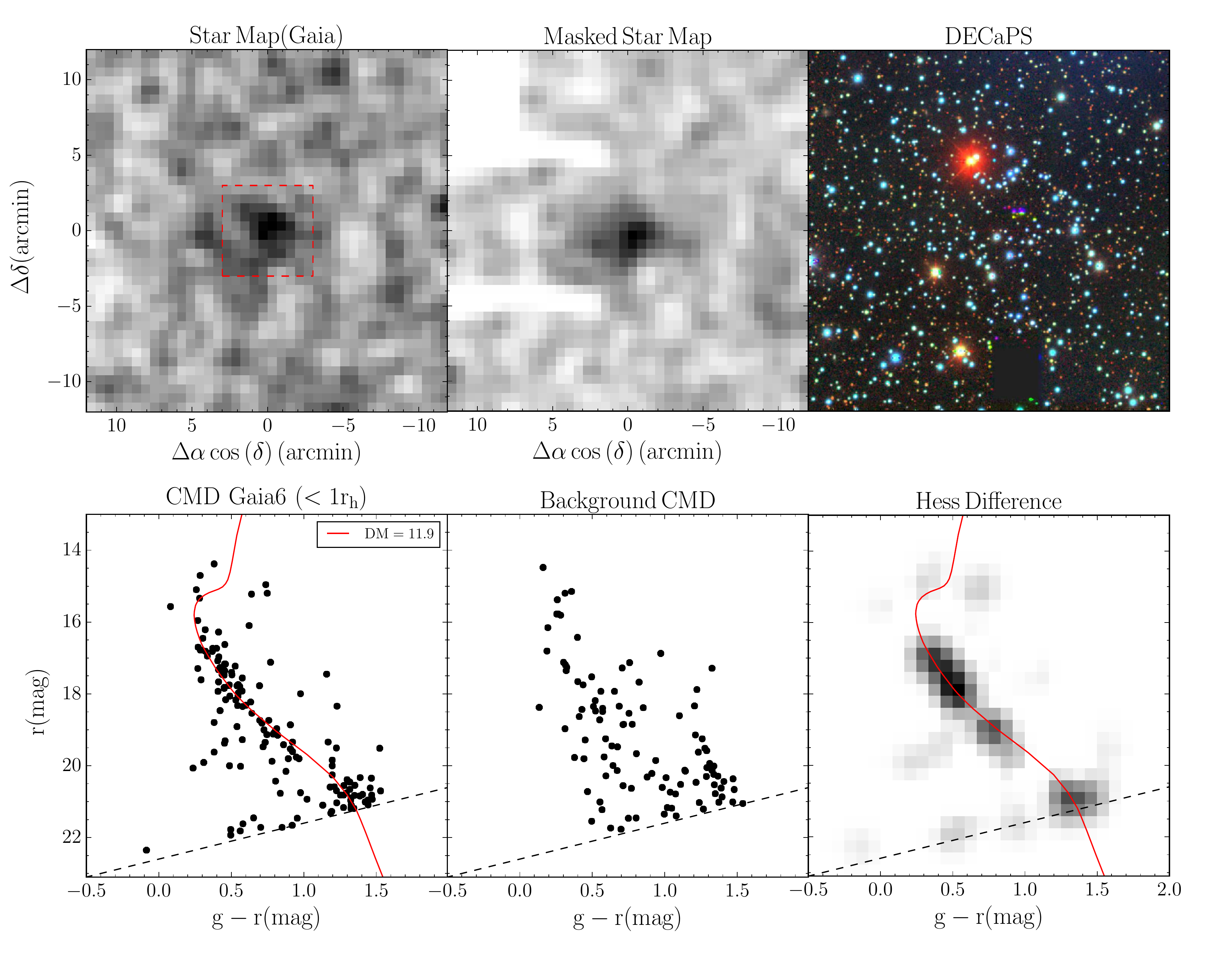}
    \caption{Discovery plot of Gaia~6. Top left panel shows a density map of Gaia stars, while the rest of the panels shows DECaPS data. All panels are described as in Figure~\ref{fig:To1}. Like Gaia~5, Gaia~6 is also within the Pan-STARRS footprint, but we only found the overndeisty in Gaia data. Noticeable features in the CMD, a clear overdensity of stars in both datasets, and hints of a distinct group of stars in the false color image, points out to Gaia~6 as a real stellar system. Gaia~6 is also close to the Galactic disk with properties that are consistent with an open Galactic cluster.}
    \label{fig:Gaia11}
\end{figure*}
%

%


\bsp	
\label{lastpage}
\end{document}